# Curvature induced modifications of chirality and magnetic configuration in perpendicular films

David Raftrey[1,2,*], Dhritiman Bhattacharya[3,*], Colin Langton[3,*], Bradley J. Fugetta[3], Subhashree Satapathy[1], Olha Bezsmertna[4], Andrea Sorrentino[5], Denys Makarov[4], Gen Yin[3], Peter Fischer[1,2,§], Kai Liu[3,#]

[1]Materials Sciences Division, Lawrence Berkeley National Laboratory, Berkeley, CA 94720, United States

[2]Department of Physics, University of California Santa Cruz, Santa Cruz, CA 95064, United States

[3]Department of Physics, Georgetown University, Washington, DC 20057, United States

[4]Helmholtz-Zentrum Dresden-Rossendorf e.V., Institute of Ion Beam Physics and Materials Research, 01328 Dresden, Germany

[5]Alba Light Source, MISTRAL beamline, 08290 Cerdanyola del Vallès, Spain

*Equal Contribution*

Email: [§] PJFischer@lbl.gov, [#] Kai.Liu@georgetown.edu

**Abstract**

Designing curvature in three-dimensional (3D) magnetic nanostructures enables controlled manipulation of local energy landscapes, allowing for the modification of noncollinear spin textures relevant for next-generation spintronic devices. In this study, we experimentally investigate 3D magnetization textures in a Co/Pd multilayer film, exhibiting strong perpendicular magnetic anisotropy (PMA), deposited onto curved Cu nanowire meshes with diameters as small as 50 nm and lengths of several microns. Utilizing magnetic soft X-ray nanotomography, we achieve reconstructions of 3D magnetic domain patterns at approximately 30 nm spatial resolution. This approach provides detailed information on both the orientation and magnitude of magnetization within the film. Our results reveal that interfacial anisotropy in the Co/Pd multilayers drives the magnetization towards the local surface normal. In contrast to typical labyrinth domains observed in planar films, the presence of curved nanowires significantly alters the domain structure, with domains preferentially aligning along the nanowire axis in close proximity, while adopting random orientations farther away. We report direct experimental observation of a curvature-induced Dzyaloshinskii-Moriya interaction (DMI), which is quantified to be approximately one-third of the intrinsic DMI in Co/Pd stacks. The curvature induced DMI enhances stability of Néel-type domain walls. These experimental observations are further supported by micromagnetic simulations. Altogether, our findings demonstrate that introducing curvature into magnetic nanostructures provides a powerful strategy for tailoring complex magnetic behaviors, paving the way for the design of advanced 3D racetrack memory and neuromorphic computing devices.

Recent advances in nanomagnetism are driving interest in 3D nanomagnetic systems, as extending nanomagnetism from 2D to 3D enables the exploration of more complex geometries, novel magnetic phenomena, e.g., 3D spin textures, and improved performance in spintronic applications. This could open up new opportunities to explore nanomagnetic systems with higher complexity, new functionalities and enhanced properties.[1-3] Significant advances in modelling and theory, synthesis and fabrication, and characterization and validation are rapidly emerging to address the challenges associated with it.[4-6]

Novel synthesis methods, such as templated and cooperative growth,[7-9] self-assembly,[10] nanoimprinting,[11] focused electron beam induced deposition (FEBID),[12-15] and two-photon-lithography[16-17] have matured and are capable of designing and fabricating tailored 3D nanostructures.[18] These nanostructures can either be directly synthesized using such methods, or indirectly by using various thin film deposition or coating methods, such as atomic layer deposition or sputtering onto 3D scaffolds.[19]

High resolution electron microscopies have been used to study the 3D stray field distribution in nanospirals,[20] topological magnetic fields in helical nanostructures,[21] 3D spin textures in skyrmion tubes,[22] vortex strings[23] or Hopfion rings.[24] Magnetic X-ray nanotomography has proven to be a powerful tool with unique features to obtain insight of 3D spin configurations, e.g. in topological spin textures such as skyrmions,[25] Bloch points,[26] vortex rings,[27] or artificially designed nanostructures, e.g. twisted nanohelices,[15] or 3D artificial spin ice systems.[28]

Curvature as a new design parameter for 3D magnetization textures has been proposed theoretically[29-30] and experimentally validated in numerous systems including rolled-up nanotubes fabricated by strain engineering [31] or on magnetically capped nano- and micron sized spheres.[32-37] The use of geometric curvature can modify magnetic interactions and local energy landscape at length scales accessible to nanofabrication and thus magnetic configurations at sub-micron length scales can be controlled.[5-6] In particular, it was theoretically predicted that geometric curvature can lead to the modification of chiral magnetic interactions with an impact on magnetization textures.[38-39] There are already first encouraging results on exploring geometric twists[40] and bends[41] for stabilization of homochiral domain walls, which stimulates further activities targeting observation of curvature induced effects on skyrmions[42] and skyrmionium[43] states.

Here, we report investigations of curvature-induced modifications of magnetic configurations in a magnetic thin film with perpendicular anisotropy deposited on an interconnected nanowire (NW) network. Interconnected magnetic NW networks are technologically interesting and have been previously studied with regard to their potential to design new types of 3D information storage and neuromorphic computing elements.[8, 44-45] In this study, the network consisting of Cu NWs with diameters of 50 nm and lengths up to several micron serves as 3D scaffolds for the synthesis of the curved film, and we focus on understanding the stabilization and modification of domain patterns as a function of the curvature governed by the NW network.

The random arrangement of NWs forming a network enables the investigation of both individual NWs with fixed curvature around their symmetry axes, as well as overlapping NWs, such as crossed configurations that introduce saddle points in the geometry. Therefore, such a NW network serves as a prototype for exploring various curvature classification categories, e.g., positive and negative curvatures. In this work, we first focus on understanding curvature-induced effects by studying a curved film above a single NW. Even in this simplified scenario, intriguing properties emerge. In the curved region, the magnetic easy axis aligns with the normal direction of the curved surface, resulting in deviations of the magnetization from the substrate normal. The demagnetization energy further drives the domains near the curved film to align parallel to the NW long axis. This results in preferential domain alignment in the film over the entire network. Additionally, we show that the chirality of the domain wall is influenced by the curvature-induced modification of the Dzyaloshinskii-Moriya interaction (DMI). We experimentally observe and quantify the strength of the curvature-induced DMI and demonstrate that it is approximately one-third of the intrinsic DMI in Co/Pd multilayers.

**Results/Discussion**

A sketch of the experimental setup of the curved platform used for magnetic soft X-ray nanotomography is shown in Fig. 1(a), consisting of NW networks coated with a multilayer thin film. The networks were prepared from electrodeposited Cu NWs with 50 nm diameter and several microns in length.[46-48] A thin film of Pd (7)/[Co (0.4)/Pd (0.6)]$_{20}$/[Co (0.7)/Pd (0.6)]$_{20}$/Ta (2.5) (all numbers in nm) was grown on the network by sputtering. Such Co/Pd multilayers exhibit strong PMA, leading to zero field stabilization of the well-known labyrinth domains with magnetization perpendicular to the substrate.[49-50] Although the multilayer structure is nominally symmetric, the presence of a 7 nm bottom Pd layer as well as strain and defects in the multilayers can introduce a net interfacial-DMI that defines the overall chirality.[51-55] The NWs are randomly dispersed, forming small, localized networks several microns in size, while the film uniformly covers the membrane, including regions with and without networks. This ensures the sample contains both flat regions and those with curvatures, enabling a detailed comparison of how curvature influences magnetic configurations.

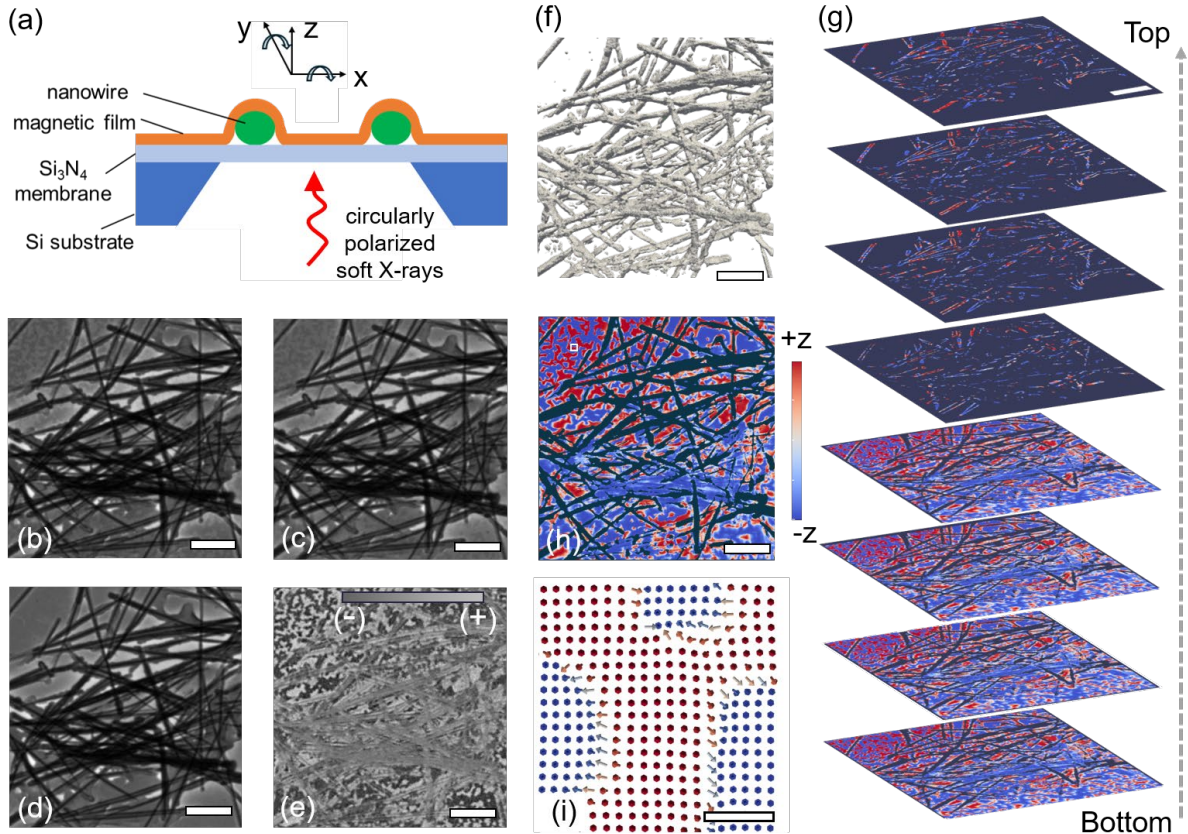

**Figure 1. 3D magnetic X-ray nanotomography of a curved thin film.** a) Schematic of experimental setup for MTXM, b) right circularly-polarized magnetic contrast ln $(T_{+\delta})$, c) left circularly-polarized magnetic contrast ln $(T_{-\delta})$, d) non-magnetic signal ln $(T_{-\delta})$+ln $(T_{+\delta})$, e) magnetic signal ln $(T_{-\delta})$-ln $(T_{+\delta})$ where positive contrast indicates magnetization pointing along the X-ray incidence direction, f) reconstructed density isosurfaces, g) reconstructed layer-by-layer magnetic information, (h) normal incidence angle of one slice from (g) with out-of-plane magnetization represented in false color, (i) zoomed-in view of magnetization vectors in the area marked by the white rectangle in panel (h). The scale bar is 1 μm in panels (b)-(h), and 45 nm in panel (i).

To visualize the 3D arrangement of magnetization, magnetic soft X-ray nanotomography was performed at the full-field soft X-ray transmission microscope at the MISTRAL beamline of the ALBA light source.[56] Figs. 1 (b, c) show the normal incidence images with right and left circular X-ray polarization, where the transmission is denoted as $T_{+\delta}$ and $T_{-\delta}$, respectively. The non-magnetic component was calculated by taking $\ln(T_{-\delta})+\ln(T_{+\delta})$ and magnetic component (dichroism) was calculated by taking $\ln(T_{-\delta})-\ln(T_{+\delta})$, as shown in Figs. 1(d) and 1(e), respectively.

To obtain 3D tomography of the magnetization, a tilt series of images was recorded with the sample tilted around the *x*-axis in the plane of the membrane (along one edge of the membrane) from –54° to +54° in 2° increments [Fig. 1(a)]. A second tilt series was then acquired after rotating the sample 90° about the *z*-axis (perpendicular to membrane), which was equivalent to tilting the sample around the *y*-axis shown in Fig. 1(a). 3D reconstructions were generated using an open-source iterative tomography solver[57-58] with an effective voxel size of ~9 nm and a half-pitch spatial resolution of ~30 nm (see Supporting Information for details). The structural density is shown in Fig. 1(f), confirming that the NW network structure is accurately captured, with the 3D stacking of wires clearly visible in the reconstruction. Fig. 1(g) presents the magnetic information across different layers of the curved film deposited onto the NW network, demonstrating the ability to reconstruct magnetization along the *z*-direction. One such layer is shown in Fig. 1(h) using false color to represent the *z*-component of the magnetization. This planar cut is in good qualitative agreement with the normal-incidence projection image [Fig. 1(e)]. Fig. 1(i) shows a zoomed-in view of the magnetic reconstruction of the area marked by the white rectangle in Fig. 1(h), where magnetization vectors illustrate the capability to resolve full 3D magnetization information. Thus, tomography provides access to spatially resolved 3D magnetic configurations within the volume. Although, the reconstruction is performed over the full 10 µm field of view, we will start in the following with a focus on the curved film above a single NW to obtain a deeper understanding of geometry induced effects in this simpler case.

**Curvature-induced variation in anisotropy direction**

In Fig. 2(a), we show the experimental full vector reconstruction of a curved section of the film delineated on a NW. Both the planar and curved regions of the film exhibit out-of-plane magnetic domains. The *x*-component of the magnetization configuration is shown in Fig. 2(b), where a pronounced difference is observed between the domain pattern in the planar and curved regions. In the planar region, the *x*-component merely demarcates the boundaries of the domains. However, in the curved region, the domains themselves exhibit a significant *x*-component with the contrast alternating between positive and negative values shown by red and blue colors, respectively. Furthermore, by examining the top view of both the *z*-component [Fig. 2(c)] and the *x*-component [Fig. 2(d)] of the magnetization in the curved region over the NW, we find that the contrast alternates within each domain along the NW. To aid visualization, the domains on the curved section are outlined with dashed lines and rectangles, with the ridge line of the NW marked with a solid line. For example, in the top-right domain 1 oriented along the +z direction, the *x*-component contrast is positive along the right side of the domain (red) and negative on the left side (blue). The adjacent domain 2, oriented along -z direction, shows the opposite behavior, with the *x*-component contrast negative on the right (blue) and positive on the left (red). This alternating contrast can be observed for most of the domains on the curved part. It demonstrates that in the curved region, the magnetization is not strictly out-of-plane within each domain as observed in the planar region, rather it fans out towards the local normal of the curved surface. This points to a possible modification of the anisotropy direction or the magnetostatic energy by the curvilinear geometry.

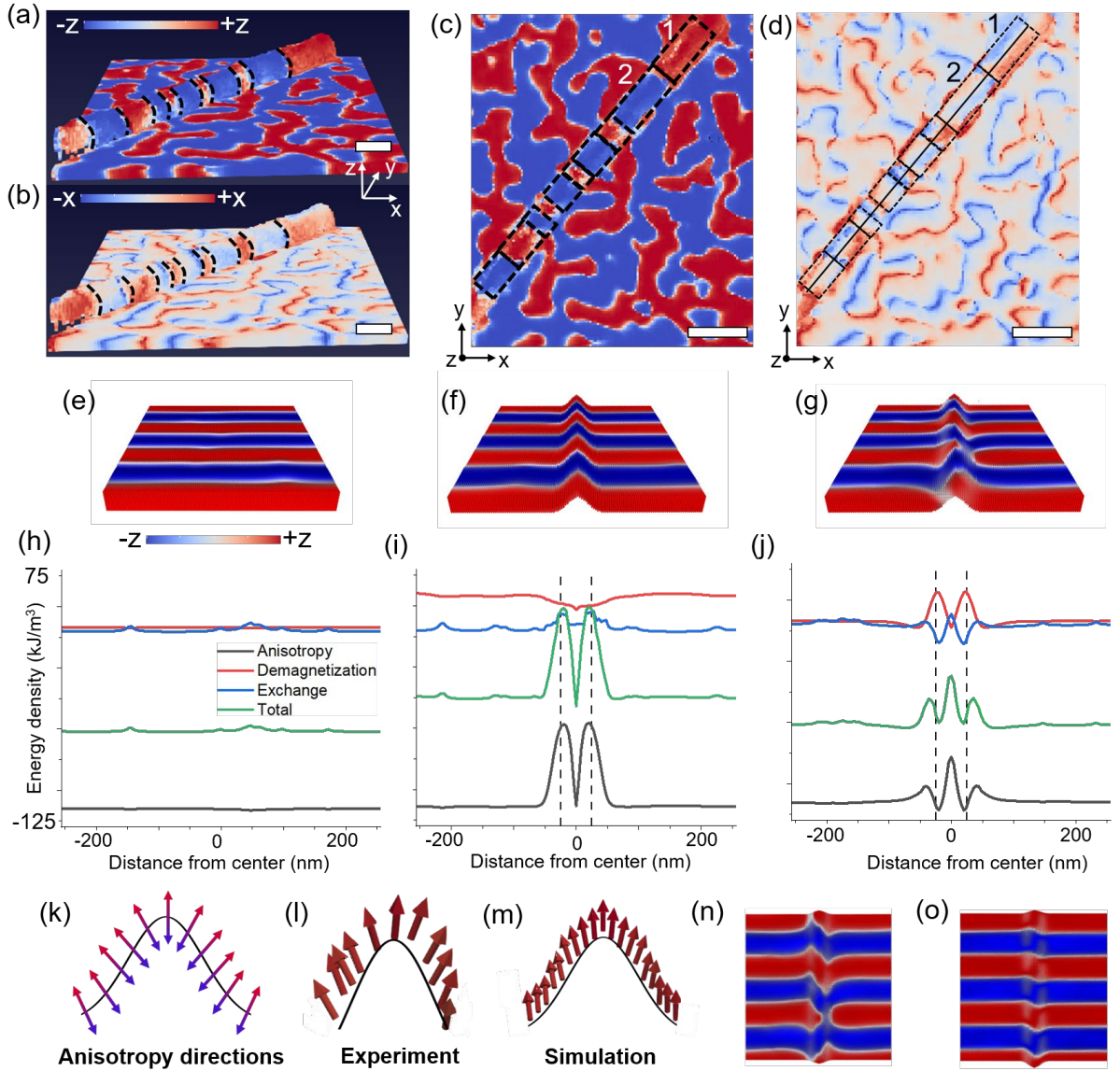

**Figure 2. Curvature induced anisotropy**. Side and top view of (a, c) the *z*-component and (b, d) the *x*-component of the magnetization in the curved film, respectively. The dashed lines and rectangles highlight the domains in the curved region. The solid line in panel (d) marks ridge line of the NW. Scale bars are 100 nm. (e-g) Micromagnetic configuration of a 512 nm × 512 nm area and (h-j) corresponding energy densities across the simulated volume for the (e, h) planar film with stripe domains, and curved film (f, i) before and (g, j) after relaxation of stripe domains, respectively. The dashed lines in panels (i, j) show the width of the Gaussian used in the simulation. (k) Schematic of anisotropy vector orientation conforming to the curved surface, (l, m) magnetization vectors along a line cut from experiment and simulation, respectively. (n, o) Top view of the micromagnetic configuration with and without considering the demagnetization energy in the simulation, respectively.

To understand this, a curved film over a single nanowire is simulated using Mumax3 with parameters typical of Co/Pd multilayers:[59] saturation magnetization $M_s$=500 kA/m, anisotropy constant $K_u$=0.15 MJ/m³, exchange stiffness $A_{ex}$=10 pJ/m.[53, 60] The total mesh size is 1020nm×1020nm×168nm with a cell size of 4nm×4nm×4nm. The thickness of the film is considered to be 52 nm. Further details of the simulation setup are provided in the Supporting Information. First, stable stripe domains with moments aligned in neighboring domains with alternating out-of-plane magnetization are formed by energy minimization in a planar film [Fig. 2(e)]. These domains are then mapped onto a curved geometry defined by a Gaussian profile, and relaxed to minimum energy state. When the magnetic easy axis is assumed to be along the ±z directions, the magnetization orientation does not rotate towards the local normal of the curved surface after relaxation [Fig. S2]. However, when the magnetic easy axis is set to be locally normal to the curved surface [Fig. 2(k)], the stripe domains in the planar regions remain largely unchanged and those on the curved surface exhibit a reduced *z*-component after relaxation, similar to the experimental observations [Fig. 2 (l, m)]. The magnetization states before and after relaxation are shown in Figs. 2(f, g). Thus, curvature induced modification of the magnetic easy axis is a necessary condition to reproduce the experimental observations. This is in line with earlier observations in Co filaments,[61] in Co/Pd and Co/Pt caps on spherical particles[33, 35] and in rolled-up architectures.[31, 62] Interestingly, the magnetization does not fully align with the new easy axis, i.e., the local normal of the curved surface [Fig. 2 (k-m)], possibly due to large exchange penalty.

Different micromagnetic energies for the three cases shown in Fig. 2 (e-g) are compared in Fig. 2 (h-j). Energies are evaluated as a function of the distance from the center of the simulated region, in the direction perpendicular to the ridge line of the curved surface. In the planar film case, the energies are constant across the film. Mapping the stripe domains onto the curved geometry leads to an increase in the total energy [Fig. 2(i)]. This can be attributed to two main factors: (1) curvature-induced variation in the magnetic easy axis - the anisotropy energy peaks at the edges of the curved region, where the magnetization deviates most from the nominal anisotropy direction, and (2) enhanced demagnetization effects. After relaxation, the anisotropy energy is decreased as the magnetization rotates to follow the easy axis [Fig. 2(j), black curve]. Furthermore, the stripe domains tend to bend away from the top of the curved surface and in some instances, they split into two branches [Fig. 2(n)]. This allows minimization of the magnetostatic energy [Fig. 2(j), red curve], which favors magnetization alignment along the long axis of the underlying NW. Simulations without the magnetostatic term does not show such alignment (Fig. 2o). Thus, due to the magnetostatic energy of the curved system, domains prefer to align parallel to the sides of the curved surface, rather than spanning directly over it. This tendency may trigger a pronounced effect on the magnetic domains in the planar region adjacent to the curved region, which we investigate next.

**Influence of curvature on domain wall orientation**

To correlate the magnetic domains with the network morphology, we explore the film over the full network beyond just a single NW and examine their alignment across the entire field of view of 10 μm. To describe and quantify the effect of curvature on the domain wall formation, in our tomographic analysis we employ the formalism of differential geometry.[63] There, a two-dimensional surface at each point is characterized by two principal curvatures $K_1$ and $K_2$. The mean curvature is the arithmetic mean of those principal curvatures $H = \frac{K_1+K_2}{2}$ and the Gaussian curvature is the square of the geometric mean of $K_1$ and $K_2$, $K = K_1 K_2$. For a nanowire geometry, one of the principal curvatures (along the wire) is zero. Details of curvature analysis is provided in the Supporting Information. With the combined magnetic and curvature maps, it is possible to investigate correlations between the images.

For this purpose, magnetic domains are binarized by selecting a threshold at $|M_z| > 0.7$, where $M_z$ is the $z$-component of the normalized magnetization, and structural features are binarized by thresholding the map of the principal curvature of the optical contrast $K_1$ at $>0.4$ [Fig. 3 (a, b)].

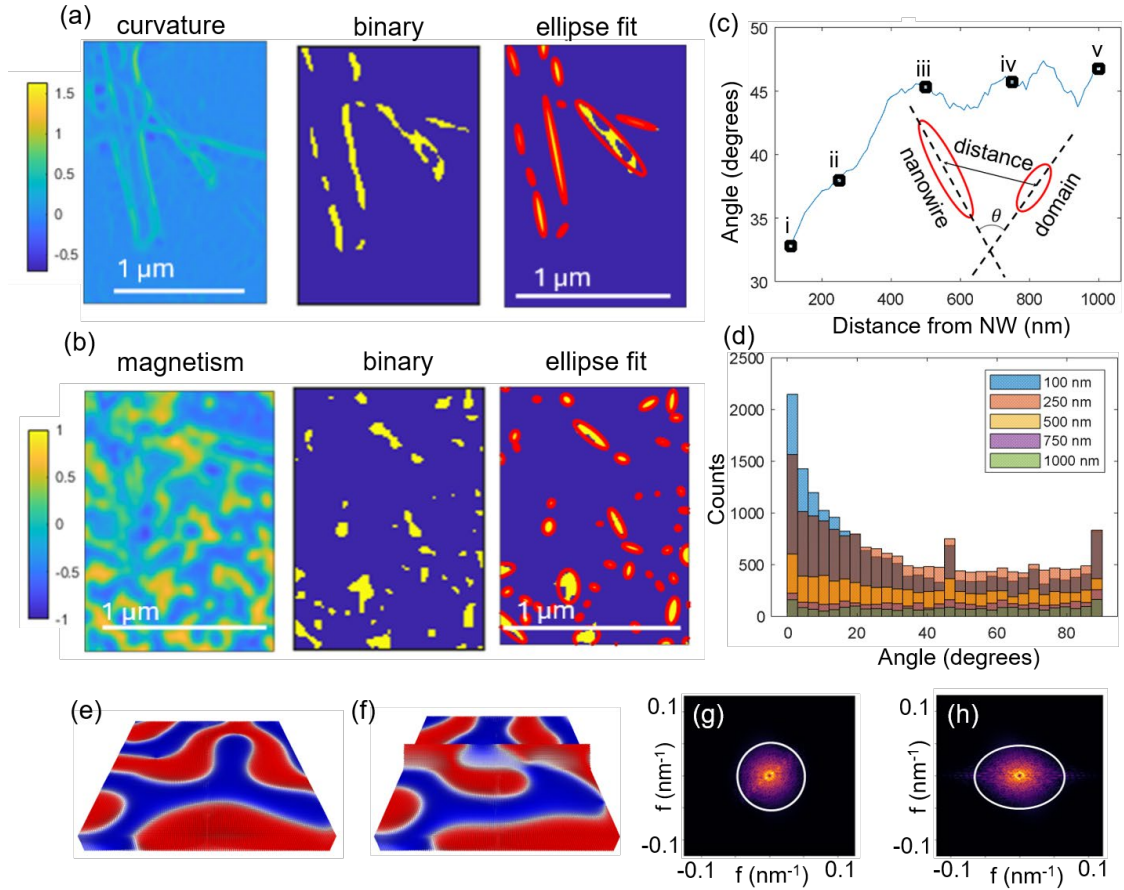

**Figure 3. Curvature induced domain alignment.** (a) Representative field of view of the map of principal curvature $K_1$, its binarized image (thresholded at $K_1 > 0.4$) and ellipse fitting of the binarized image. (b) Same region in magnetic domains, binary segmentation selecting greatest magnetic contrast and ellipse fitting of regions of greatest magnetic contrast, where the magnetic domains are selected as the region with $|M_z| > 0.7$. (c) Ellipse fitting of domains at different distances from NWs, from 100 nm to 1000 nm (i-v). (d) Plot of mean from histograms showing average angle between NW and magnetic domain as a function of separation corresponding to the distances labeled (i-v) in (c), converging near 45° beyond 400 nm separation. (e, f) Micromagnetic configuration in a 512 nm×512 nm area and (g, h) Fourier transform of the domain pattern in the planar and the curved geometry, respectively.

The map of principal curvatures is calculated based on the analysis of the transmitted intensity as discussed in Supporting Information Section SI5 [Fig. S4]. From the binary maps, ellipse fitting extracts the orientation, positions and size of the magnetic and curvature regions. The orientation of the domain walls and the NWs can be quantified as the angle between the major axes of the fitting ellipses. We find that at distances close to the NWs, the domains are oriented more along the axis of the wire (e.g. an average angle ~33° for 100 nm), and at distances far from the NW (1000 nm), the domains are oriented at random angles with an average angle of ~45° [Fig. 3(c)]. The angle between NW and domain is taken for each domain as the angle with the nearest NW. The statistics are collected for all rotation angles. For each radial distance the domains are binned into 100 nm wide segments to define a population of domains. For each population of domains, a histogram is generated [Fig. 3 (d)]. The histograms are skewed towards 0°, which is the parallel orientation of domain and nanowire. For larger distances the average angle converges to 45° which is expected for a population of randomly oriented line segments.

Micromagnetic simulations are performed to verify these effects using labyrinth domains, as observed experimentally. Figure 3(e) shows that a typical labyrinth domain pattern stabilizes in a planar film upon relaxation from a random magnetization state. Using the same random seed and introducing curvature, a clear difference is found both on the curved region and in the adjacent planar areas [Fig. 3(f)]. For example, the orientation of the central domain remains random in the planar region, but on the curved region, it expands along the ridge line. Additionally, this reorientation influences neighboring domains; the domain at the edge becomes more aligned with the ridge line of the curved surface [Fig. 3(f)]. These qualitative observations are further examined using Fast Fourier Transform (FFT) analysis. Simulating domain relaxation in planar and curved films across 20 different random magnetization seeds reveals a consistent trend. For the planar film, a ring structure in the FFT [Fig. 3(g)] indicates uniform domain periodicity with random orientations. In contrast, when the domains from the curved structure are mapped onto a 2D plane preserving spatial relationships for the FFT analysis, a distinct alignment can be seen along the ridge line, evident from the ovular shape in Fig. 3(h). Hence, the geometry induces domain realignment by guiding the magnetization to follow the principal direction with zero curvature, leading to preferred orientations along the curved contours. Micromagnetic simulation shows that the magnetization reversal under external field is also strongly influenced by such curvature induced domain alignment (Fig. S3).

**Impact of curvature on chirality of domain walls**

Lastly, the effect of curvature on the chirality of domain walls (DWs) is examined. The DW chirality is characterized by measuring the angle between the DW magnetization (**m**) and DW normal vector (**n**) [Fig. 4(a) inset].[52] The area shown in Fig. 2(a) is analyzed, consisting of 120 pixels × 170 pixels with dimensions 1080 nm × 1530 nm. In the planar region, a predominance of right-handed Néel walls is observed as shown in the histogram [Fig. 4(a)], a manifestation of the DMI in the system. However, a significant number of DWs deviate from the purely right-handed Néel type. In contrast, in the curved part of the film, right-handed Néel DWs are more prevalent, as indicated by the tighter distribution in the histogram [Fig. 4(b)]. Magnetization profile of two DWs, one from the top and one from the side of the curved region are shown in Fig. 4(c). For the DW located at the top, the magnetic easy axis is completely out-of-plane while demagnetization energy favors in-plane alignment of magnetic moments parallel to the long axis of the NW (now defined as the *x*-axis in Fig. 4). As a result, a right-handed Néel DW is stabilized and characterized by a strong positive *x*-component and minimal *y*-component of magnetization (in the base plane and orthogonal to *x* direction) while the normalized *z*-component switches between ±1 [Fig. 4(d)]. At the sides of the curved region, a more complex behavior seems to emerge. As discussed earlier, magnetization in these regions tends to fan out to align with the local surface normal. This gives rise to domain walls with distinct characteristics, where the DW exhibits a sizeable *y*-component and a reduced *z*-component contrast, transitioning between approximately ±0.75 rather than fully switching between ±1 [Fig. 4(e)]. This resembles a twisting of the domain wall structure. On the other side of the curved surface, the twisting seems to be in the opposite direction with a reversed *y*-component [Fig. S5]. However, if expressed in a coordinate basis rotated in the *yz* plane to align with the direction of maximum $|M_z|$, represented by $x_R$-$y_R$-$z_R$ in Fig. 4(c), these walls also correspond to Néel DWs (Fig. 4f).

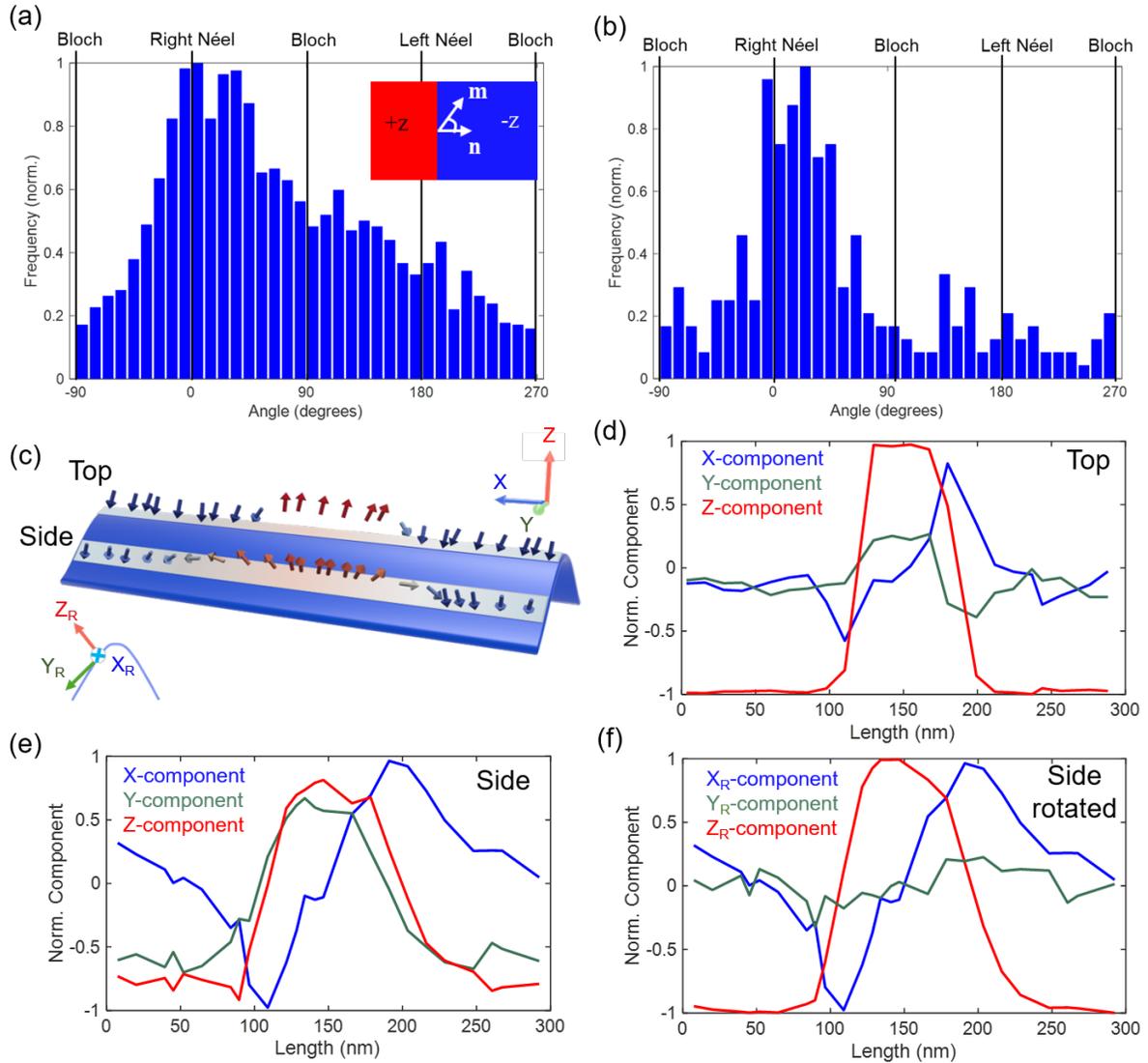

**Figure 4. Curvature modified chirality of domain walls.** Histograms of angle between DW magnetization and DW normal vector, measured pixel-by-pixel along the DW centerline in (a) the planar and (b) the curved region of the film, respectively. (c) Schematic illustration of the experimentally observed magnetic configurations (d) on top and (e,f) side of the curved region, in (d,e) the $x$-$y$-$z$ and (f) $x_R$-$y_R$-$z_R$ coordinates.

Therefore, curvature strongly influences domain wall chirality, promoting Néel-type walls. This finding is indicative of the additional contribution stemming from the curvature induced DMI, which adds to the intrinsic interfacial DMI in Co/Pd stacks. From the histograms, we estimate the fraction of right-handed Néel DWs as [($\sum_{-45°}^{+45°} counts$ / $\sum_{-90°}^{+270°} counts$)×100%], where 0° corresponds to purely right-handed Néel DWs. This fraction is found to be 40.8% in the planar region and 53.9% in the curved region. This 13.1% enhancement suggests that curvature contributes an effective DMI approximately one-third the strength of the intrinsic DMI in Co/Pd stacks. It can be attributed to the magnetostatic energy which favors alignment parallel to the long NW axis similar to cylindrical magnetic NWs. This gives rise to the strong magnetization component along that direction and consequently more right-handed Néel DWs are stabilized. Theoretically, such curvature induced DMI is calculated as $D_c = 2A \times$ curvature,[30] where curvature for the case of a cylindrical surface is $1/R$, with $R$ the radius of cylinder and $A$ the exchange constant. For the geometry of our NWs ($R = 25$ nm) and assuming $A = 10^{-11}$ J/m, we obtain $D_c = 0.8$ mJ/m$^2$. Additionally, strain gradients induced by curvature and increased roughness from the NWs could also enhance DMI.[64-66]

DMI in Co/Pd multilayers have been measured previously. Even in symmetric multilayers, significant DMI can arise due to unequal strain from the bottom Pd/Co and top Co/Pd interfaces.[54] For a single Pd/Co interface, DMI is typically negative.[51-52] The sign of DMI in multilayers depend on Pd layer thickness,[54] and in trilayers it was reported to vary between ±0.3 mJ/m$^2$. In our case, the sign of DMI was positive as indicated by the right-handed Néel DWs. The DMI strength also strongly depends on the number of Co/Pd bilayers and has been shown to increase threefold (from ~1 mJ/m$^2$ to ~3 mJ/m$^2$) when repetition number increases from 1 to 20. Therefore, a significant intrinsic DMI of up to 3 mJ/m$^2$ can be expected in our system, and the contribution from the curvature-induced DMI estimated at 0.8 mJ/m$^2$, would be roughly one-third of the intrinsic DMI as assessed based on the analysis of the histograms in Fig. 4. To our knowledge, this is the first experimental quantification of the curvature induced DMI in perpendicularly magnetized curved thin films. These results will motivate future experiments to quantify DMI in curved magnetic films using alternative methods such as asymmetric domain wall propagation and Brillouin light scattering.[67] Furthermore, in contrast to prior observations in in-plane systems,[41] our results pave the way to experimental search of curvature stabilized topological spin textures.[42] These findings will also open new avenues for 3D curved racetrack memory.[3, 40]

**Conclusions**

In summary, we have investigated the effect of curvature on magnetic configurations in a Co/Pd multilayer film with PMA coated on a NW network scaffold. The sample was measured with soft X-rays with circular polarization for magnetic contrast. The data were reconstructed in 3D using vector-resolved nanotomography and the domain structure was found to be continuous across the planar and curved parts of the sample. The curved geometry modifies the orientation of the easy axis and the magnetization follows the local normal of the curved surface. The correlation between domain structure and curvature is quantified by measuring the orientation of the domains in the sample as a function of separation to the nearest NW which shows parallel alignment at small distances and random alignment at larger distances. Furthermore, our work experimentally confirms the theoretical concept of curvature induced DMI in PMA thin films and paves the way towards stabilization of curvature induced skyrmions and skyrmionium states. Micromagnetic simulations support our experimental findings, confirming that the energy landscape is modified on the curved surface. This work demonstrates that curvature modifies the local magnetic easy axis, domain structure and chirality of domain walls, offering an effective method to tailor spin textures and energy landscapes at the nanoscale.

**Methods/Experimental**

**Sample Fabrication**

Cu NWs with 50 nm diameter and several microns in length were first prepared by electrodeposition into nanoporous polycarbonate membranes. The NWs were harvested by dissolving the membrane in dichloromethane, then liquid-exchanged with deionized water before being drop-cast onto a 50nm thin silicon nitride ($Si_3N_4$) membrane. The $Si_3N_4$ membrane allows for sufficient transparency in the soft X-ray regime required to perform magnetic transmission X-ray microscopy (MTXM). The Co/Pd film was grown on the network by DC magnetron sputtering under a base pressure below $3\times10^{-8}$ Torr and an Ar working pressure of 5 mTorr.

**Magnetic X-Ray nanotomography**

For magnetic soft X-ray nanotomography, images were recorded at the Co $L_3$ edge with both left and right polarization, divided by the incident intensity to obtain the transmission, and subtracted from each other to reduce the non-magnetic background in the magnetic images. To obtain 3D tomography of the magnetization, a tilt series of images was recorded and reconstructed using an iterative algorithm. The input to the reconstruction algorithm is an aligned and normalized tilt series containing both magnetic and non-magnetic information. Using the aligned image stack, the algorithm reconstructs the structural and magnetic information in a 3D volume.

**Supporting Information**

Details of reconstruction, micromagnetic simulation setup, micromagnetic simulation without anisotropy direction modification, micromagnetic simulation with applied field and temperature, curvature analysis and curvature induced chirality.

**Associated Content**

A pre-print version of this manuscript has been posted on the ArXiv preprint server as David Raftrey; Dhritiman Bhattacharya; Colin Langton; Bradley Fugetta; S. Satapathy; Olha Bezsmertna; Andrea Sorrentino; Denys Makarov; Gen Yin; Peter Fischer; Kai Liu. Curvature induced modifications of chirality and magnetic configuration in perpendicular magnetized films. 2025, Article number 2506.05938. ArXiv. https://arxiv.org/abs/2506.05938.

**Acknowledgements**

This work was primarily funded by the U.S. Department of Energy, Office of Science, Office of Basic Energy Sciences, Materials Sciences and Engineering Division under Contract No. DE-AC02-05-CH11231 (NEMM program MSMAG). Work at GU (sample synthesis, analysis and micromagnetic simulations) was supported by the NSF (DMR-2005108 and ECCS-2429995). The MTXM experiments were performed at the MISTRAL beamline at ALBA light source in collaboration with ALBA staff. The ALBA light source is funded by the Ministry of Research and Innovation of Spain, by the Generalitat de Catalunya and by European FEDER funds. Work at the Molecular Foundry was supported by the U.S. Department of Energy, Office of Science, Office of Basic Energy Sciences, of the U.S. Department of Energy under Contract No. DE-AC02-05CH11231. Work at HZDR was supported in part via German Research Foundation (grants MA5144/22-1, MA5144/33-1) and ERC grant 3DmultiFerro (Project number: 101141331). We thank Dr. Oleksandr Pylypovskyi (HZDR) for fruitful discussions.

# References

1. Fernández-Pacheco, A.; Streubel, R.; Fruchart, O.; Hertel, R.; Fischer, P.; Cowburn, R. P., Three-dimensional nanomagnetism. *Nature Communications* **2017**, *8*, 15756.
2. Fischer, P.; Sanz-Hernández, D.; Streubel, R.; Fernández-Pacheco, A., Launching a new dimension with 3D magnetic nanostructures. *APL Materials* **2020**, *8* (1), 010701.
3. Gu, K.; Guan, Y.; Hazra, B. K.; Deniz, H.; Migliorini, A.; Zhang, W.; Parkin, S. S. P., Three-dimensional racetrack memory devices designed from freestanding magnetic heterostructures. *Nature Nanotechnology* **2022**, *17* (10), 1065-1071.
4. Gentile, P.; Cuoco, M.; Volkov, O. M.; Ying, Z.-J.; Vera-Marun, I. J.; Makarov, D.; Ortix, C., Electronic materials with nanoscale curved geometries. *Nature Electronics* **2022**, *5* (9), 551-563.
5. Makarov, D.; Volkov, O. M.; Kákay, A.; Pylypovskyi, O. V.; Budinská, B.; Dobrovolskiy, O. V., New Dimension in Magnetism and Superconductivity: 3D and Curvilinear Nanoarchitectures. *Advanced Materials* **2022**, *34* (3), 2101758.
6. Sheka, D. D.; Pylypovskyi, O. V.; Volkov, O. M.; Yershov, K. V.; Kravchuk, V. P.; Makarov, D., Fundamentals of Curvilinear Ferromagnetism: Statics and Dynamics of Geometrically Curved Wires and Narrow Ribbons. *Small* **2022**, *18* (12), 2105219.
7. Burks, E. C.; Gilbert, D. A.; Murray, P. D.; Flores, C.; Felter, T. E.; Charnvanichborikarn, S.; Kucheyev, S. O.; Colvin, J. D.; Yin, G.; Liu, K., 3D Nanomagnetism in Low Density Interconnected Nanowire Networks. *Nano Lett* **2021**, *21* (1), 716-722.
8. Bhattacharya, D.; Chen, Z.; Jensen, C. J.; Liu, C.; Burks, E. C.; Gilbert, D. A.; Zhang, X.; Yin, G.; Liu, K., 3D Interconnected Magnetic Nanowire Networks as Potential Integrated Multistate Memristors. *Nano Letters* **2022**, *22* (24), 10010-10017.
9. Chen, F.; Yang, Z.; Li, J.-N.; Jia, F.; Wang, F.; Zhao, D.; Peng, R.-W.; Wang, M., Formation of magnetic nanowire arrays by cooperative lateral growth. *Sci. Adv.* **2022**, *8* (4), eabk0180.
10. Ahn, J.; Ha, J.-H.; Jeong, Y.; Jung, Y.; Choi, J.; Gu, J.; Hwang, S. H.; Kang, M.; Ko, J.; Cho, S.; Han, H.; Kang, K.; Park, J.; Jeon, S.; Jeong, J.-H.; Park, I., Nanoscale three-dimensional fabrication based on mechanically guided assembly. *Nature Communications* **2023**, *14* (1), 833.
11. Wu, H.; Tian, Y.; Luo, H.; Zhu, H.; Duan, Y.; Huang, Y., Fabrication Techniques for Curved Electronics on Arbitrary Surfaces. *Advanced Materials Technologies* **2020**, *5* (8), 2000093.
12. Pablo-Navarro, J.; Sanz-Hernández, D.; Magén, C.; Fernández-Pacheco, A.; de Teresa, J. M., Tuning shape, composition and magnetization of 3D cobalt nanowires grown by focused electron beam induced deposition (FEBID). *Journal of Physics D: Applied Physics* **2017**, *50* (18), 18LT01.
13. Huth, M.; Porrati, F.; Schwalb, C.; Winhold, M.; Sachser, R.; Dukic, M.; Adams, J.; Fantner, G., Focused electron beam induced deposition: A perspective. *Beilstein Journal of Nanotechnology* **2012**, *3*, 597-619.
14. Winkler, R.; Fowlkes, J. D.; Rack, P. D.; Plank, H., 3D nanoprinting via focused electron beams. *Journal of Applied Physics* **2019**, *125* (21), 210901.
15. Sanz-Hernández, D.; Hierro-Rodriguez, A.; Donnelly, C.; Pablo-Navarro, J.; Sorrentino, A.; Pereiro, E.; Magén, C.; McVitie, S.; de Teresa, J. M.; Ferrer, S.; Fischer, P.; Fernández-Pacheco, A., Artificial Double-Helix for Geometrical Control of Magnetic Chirality. *ACS Nano* **2020**, *14* (7), 8084-8092.
16. Harinarayana, V.; Shin, Y. C., Two-photon lithography for three-dimensional fabrication in micro/nanoscale regime: A comprehensive review. *Optics & Laser Technology* **2021**, *142*, 107180.
17. Xiong, X.; Jiang, S.-C.; Hu, Y.-H.; Peng, R.-W.; Wang, M., Structured Metal Film as a Perfect Absorber. *Adv. Mater.* **2013**, *25* (29), 3994-4000.
18. Huang, Y.; Wu, H.; Xiao, L.; Duan, Y.; Zhu, H.; Bian, J.; Ye, D.; Yin, Z., Assembly and applications of 3D conformal electronics on curvilinear surfaces. *Materials Horizons* **2019**, *6* (4), 642-683.
19. Bezsmertna, O.; Xu, R.; Pylypovskyi, O.; Raftrey, D.; Sorrentino, A.; Fernandez-Roldan, J. A.; Soldatov, I.; Wolf, D.; Lubk, A.; Schäfer, R.; Fischer, P.; Makarov, D., Magnetic Solitons in Hierarchical 3D Magnetic Nanoarchitectures of Nanoflower Shape. *Nano Letters* **2024**, *24* (49), 15774-15780.

20. Phatak C, L. Y. G. E. B. S. D. S. E.; Petford-Long, A., Magnetic structure of 3D sculpted cobalt nanoparticles. *Nano Lett.* **2014,** *14*, 759.
21. Fullerton, J.; Phatak, C., Design and Control of Three-Dimensional Topological Magnetic Fields Using Interwoven Helical Nanostructures. *Nano Letters* **2025,** *25* (13), 5148-5155.
22. Wolf, D.; Schneider, S.; Rößler, U. K.; Kovács, A.; Schmidt, M.; Dunin-Borkowski, R. E.; Büchner, B.; Rellinghaus, B.; Lubk, A., Unveiling the three-dimensional magnetic texture of skyrmion tubes. *Nature Nanotechnology* **2022,** *17* (3), 250-255.
23. Volkov, O. M.; Wolf, D.; Pylypovskyi, O. V.; Kákay, A.; Sheka, D. D.; Büchner, B.; Fassbender, J.; Lubk, A.; Makarov, D., Chirality coupling in topological magnetic textures with multiple magnetochiral parameters. *Nature Communications* **2023,** *14* (1), 1491.
24. Zheng, F.; Kiselev, N. S.; Rybakov, F. N.; Yang, L.; Shi, W.; Blügel, S.; Dunin-Borkowski, R. E., Hopfion rings in a cubic chiral magnet. *Nature* **2023,** *623* (7988), 718-723.
25. Raftrey, D.; Finizio, S.; Chopdekar, R. V.; Dhuey, S.; Bayaraa, T.; Ashby, P.; Raabe, J.; Santos, T.; Griffin, S.; Fischer, P., Quantifying the topology of magnetic skyrmions in three dimensions. *Science Advances* **2024,** *10* (40), eadp8615.
26. Donnelly, C.; Guizar-Sicairos, M.; Scagnoli, V.; Gliga, S.; Holler, M.; Raabe, J.; Heyderman, L. J., Three-dimensional magnetization structures revealed with X-ray vector nanotomography. *Nature* **2017,** *547*, 328.
27. Donnelly, C.; Metlov, K. L.; Scagnoli, V.; Guizar-Sicairos, M.; Holler, M.; Bingham, N. S.; Raabe, J.; Heyderman, L. J.; Cooper, N. R.; Gliga, S., Experimental observation of vortex rings in a bulk magnet. *Nature Physics* **2021,** *17* (3), 316-321.
28. Harding, E.; Araki, T.; Askey, J.; Hunt, M.; Van Den Berg, A.; Raftrey, D.; Aballe, L.; Kaulich, B.; MacDonald, E.; Fischer, P.; Ladak, S., Imaging the magnetic nanowire cross section and magnetic ordering within a suspended 3D artificial spin-ice. *APL Materials* **2024,** *12* (2), 021116.
29. Gaididei, Y.; Kravchuk, V. P.; Sheka, D. D., Curvature Effects in Thin Magnetic Shells. *Physical Review Letters* **2014,** *112* (25), 257203.
30. Sheka, D. D.; Pylypovskyi, O. V.; Landeros, P.; Gaididei, Y.; Kákay, A.; Makarov, D., Nonlocal chiral symmetry breaking in curvilinear magnetic shells. *Communications Physics* **2020,** *3* (1), 128.
31. Streubel, R.; Kronast, F.; Fischer, P.; Parkinson, D.; Schmidt, O. G.; Makarov, D., Retrieving spin textures on curved magnetic thin films with full-field soft X-ray microscopies. *Nature Communications* **2015,** *6* (1), 7612.
32. Streubel, R.; Kronast, F.; Reiche, C. F.; Mühl, T.; Wolter, A. U. B.; Schmidt, O. G.; Makarov, D., Vortex circulation and polarity patterns in closely packed cap arrays. *Applied Physics Letters* **2016,** *108* (4), 042407.
33. Ulbrich, T. C.; Bran, C.; Makarov, D.; Hellwig, O.; Risner-Jamtgaard, J. D.; Yaney, D.; Rohrmann, H.; Neu, V.; Albrecht, M., Effect of magnetic coupling on the magnetization reversal in arrays of magnetic nanocaps. *Physical Review B* **2010,** *81* (5), 054421.
34. Makarov, D.; Baraban, L.; Guhr, I. L.; Boneberg, J.; Schift, H.; Gobrecht, J.; Schatz, G.; Leiderer, P.; Albrecht, M., Arrays of magnetic nanoindentations with perpendicular anisotropy. *Applied Physics Letters* **2007,** *90* (9), 093117.
35. Ulbrich, T. C.; Makarov, D.; Hu, G.; Guhr, I. L.; Suess, D.; Schrefl, T.; Albrecht, M., Magnetization Reversal in a Novel Gradient Nanomaterial. *Physical Review Letters* **2006,** *96* (7), 077202.
36. Baraban, L.; Makarov, D.; Albrecht, M.; Rivier, N.; Leiderer, P.; Erbe, A., Frustration-induced magic number clusters of colloidal magnetic particles. *Physical Review E* **2008,** *77* (3), 031407.
37. Brandt, R.; Ruumlckriem, R.; Gilbert, D. A.; Ganss, F.; Senn, T.; Liu, K.; Albrecht, M.; Schmidt, H., Size-dependent magnetization switching characteristics and spin wave modes of FePt nanostructures. *J. Appl. Phys.* **2013,** *113* (20), 203910.
38. Volkov, O. M.; Sheka, D. D.; Gaididei, Y.; Kravchuk, V. P.; Rößler, U. K.; Fassbender, J.; Makarov, D., Mesoscale Dzyaloshinskii-Moriya interaction: geometrical tailoring of the magnetochirality. *Scientific Reports* **2018,** *8* (1), 866.

39. Kravchuk, V. P.; Rößler, U. K.; Volkov, O. M.; Sheka, D. D.; van den Brink, J.; Makarov, D.; Fuchs, H.; Fangohr, H.; Gaididei, Y., Topologically stable magnetization states on a spherical shell: Curvature-stabilized skyrmions. *Physical Review B* **2016,** *94* (14), 144402.
40. Farinha, A. M. A.; Yang, S.-H.; Yoon, J.; Pal, B.; Parkin, S. S. P., Interplay of geometrical and spin chiralities in 3D twisted magnetic ribbons. *Nature* **2025,** *639* (8053), 67-72.
41. Volkov, O. M.; Kákay, A.; Kronast, F.; Mönch, I.; Mawass, M.-A.; Fassbender, J.; Makarov, D., Experimental Observation of Exchange-Driven Chiral Effects in Curvilinear Magnetism. *Physical Review Letters* **2019,** *123* (7), 077201.
42. Kravchuk, V. P.; Sheka, D. D.; Kákay, A.; Volkov, O. M.; Rößler, U. K.; van den Brink, J.; Makarov, D.; Gaididei, Y., Multiplet of Skyrmion States on a Curvilinear Defect: Reconfigurable Skyrmion Lattices. *Physical Review Letters* **2018,** *120* (6), 067201.
43. Pylypovskyi, O. V.; Makarov, D.; Kravchuk, V. P.; Gaididei, Y.; Saxena, A.; Sheka, D. D., Chiral Skyrmion and Skyrmionium States Engineered by the Gradient of Curvature. *Physical Review Applied* **2018,** *10* (6), 064057.
44. Araujo, E.; Encinas, A.; Velázquez-Galván, Y.; Martínez-Huerta, J. M.; Hamoir, G.; Ferain, E.; Piraux, L., Artificially modified magnetic anisotropy in interconnected nanowire networks. *Nanoscale* **2015,** *7* (4), 1485-1490.
45. Bhattacharya, D.; Langton, C.; Rajib, M. M.; Marlowe, E.; Chen, Z.; Al Misba, W.; Atulasimha, J.; Zhang, X.; Yin, G.; Liu, K., Self-assembled 3D Interconnected Magnetic Nanowire Networks for Neuromorphic Computing. *ACS Appl. Mater. Interfaces* **2025,** *17*, 20087-20095.
46. Liu, K.; Nagodawithana, K.; Searson, P. C.; Chien, C. L., Perpendicular giant magnetoresistance of multilayered Co/Cu nanowires. *Phys. Rev. B* **1995,** *51* (11), 7381.
47. Hong, K.; Yang, F. Y.; Liu, K.; Reich, D. H.; Searson, P. C.; Chien, C. L.; Balakirev, F. F.; Boebinger, G. S., Giant positive magnetoresistance of Bi nanowire arrays in high magnetic fields. *J. Appl. Phys.* **1999,** *85*, 6184.
48. Malloy, J.; Quintana, A.; Jensen, C. J.; Liu, K., Efficient and Robust Metallic Nanowire Foams for Deep Submicrometer Particulate Filtration. *Nano Lett.* **2021,** *21* (7), 2968-2974.
49. Gilbert, D. A.; Maranville, B. B.; Balk, A. L.; Kirby, B. J.; Fischer, P.; Pierce, D. T.; Unguris, J.; Borchers, J. A.; Liu, K., Realization of Ground State Artificial Skyrmion Lattices at Room Temperature. *Nat. Commun.* **2015,** *6*, 8462.
50. Davies, J. E.; Hellwig, O.; Fullerton, E. E.; Denbeaux, G.; Kortright, J. B.; Liu, K., Magnetization reversal of Co/Pt multilayers: Microscopic origin of high-field magnetic irreversibility. *Phys. Rev. B* **2004,** *70* (22), 224434.
51. Chen, G.; Ophus, C.; Lo Conte, R.; Wiesendanger, R.; Yin, G.; Schmid, A. K.; Liu, K., Ultrasensitive Sub-monolayer Palladium Induced Chirality Switching and Topological Evolution of Skyrmions. *Nano Lett.* **2022,** *22*, 6678-6684.
52. Chen, G.; Mascaraque, A.; Jia, H.; Zimmermann, B.; Robertson, M.; Conte, R. L.; Hoffmann, M.; González Barrio, M. A.; Ding, H.; Wiesendanger, R.; Michel, E. G.; Blügel, S.; Schmid, A. K.; Liu, K., Large Dzyaloshinskii-Moriya Interaction Induced by Chemisorbed Oxygen on a Ferromagnet Surface. *Sci. Adv.* **2020,** *6* (33), eaba4924.
53. Pollard, S. D.; Garlow, J. A.; Yu, J.; Wang, Z.; Zhu, Y.; Yang, H., Observation of stable Néel skyrmions in cobalt/palladium multilayers with Lorentz transmission electron microscopy. *Nature Communications* **2017,** *8* (1), 14761.
54. Davydenko, A. V.; Kozlov, A. G.; Stebliy, M. E.; Kolesnikov, A. G.; Sarnavskiy, N. I.; Iliushin, I. G.; Golikov, A. P., Dzyaloshinskii-Moriya interaction and chiral damping effect in symmetric epitaxial Pd/Co/Pd(111) trilayers. *Physical Review B* **2021,** *103* (9), 094435.
55. Davydenko, A. V.; Kozlov, A. G.; Kolesnikov, A. G.; Stebliy, M. E.; Suslin, G. S.; Vekovshinin, Y. E.; Sadovnikov, A. V.; Nikitov, S. A., Dzyaloshinskii-Moriya interaction in symmetric epitaxial [Co/Pd(111)]$_N$ superlattices with different numbers of Co/Pd bilayers. *Physical Review B* **2019,** *99* (1), 014433.

56. Sorrentino, A.; Nicolas, J.; Valcarcel, R.; Chichon, F. J.; Rosanes, M.; Avila, J.; Tkachuk, A.; Irwin, J.; Ferrer, S.; Pereiro, E., MISTRAL: a transmission soft X-ray microscopy beamline for cryo nano-tomography of biological samples and magnetic domains imaging. *Journal of Synchrotron Radiation* **2015,** *22* (4), 1112-1117.
57. Donnelly, C.; Gliga, S.; Scagnoli, V.; Holler, M.; Raabe, J.; Heyderman, L. J.; Guizar-Sicairos, M., Tomographic reconstruction of a three-dimensional magnetization vector field. *New Journal of Physics* **2018,** *20* (8), 083009.
58. Schwartz, J.; Harris, C.; Pietryga, J.; Zheng, H.; Kumar, P.; Visheratina, A.; Kotov, N. A.; Major, B.; Avery, P.; Ercius, P.; Ayachit, U.; Geveci, B.; Muller, D. A.; Genova, A.; Jiang, Y.; Hanwell, M.; Hovden, R., Real-time 3D analysis during electron tomography using tomviz. *Nature Communications* **2022,** *13* (1), 4458.
59. Vansteenkiste, A.; Leliaert, J.; Dvornik, M.; Helsen, M.; Garcia-Sanchez, F.; Van Waeyenberge, B., The design and verification of MuMax3. *AIP Advances* **2014,** *4* (10), 107133.
60. Greene, P. K.; Kirby, B. J.; Lau, J. W.; Borchers, J. A.; Fitzsimmons, M. R.; Liu, K., Deposition order dependent magnetization reversal in pressure graded Co/Pd films. *Applied Physics Letters* **2014,** *104* (15), 152401.
61. Chen, F.; Wang, F.; Jia, F.; Li, J.; Liu, K.; Huang, S.; Luan, Z.; Wu, D.; Chen, Y.; Zhu, J.; Peng, R. W.; Wang, M., Periodic spin configuration in single-crystalline cobalt filaments. *Phys. Rev. B* **2016,** *93*, 054405.
62. Smith, E. J.; Makarov, D.; Sanchez, S.; Fomin, V. M.; Schmidt, O. G., Magnetic Microhelix Coil Structures. *Physical Review Letters* **2011,** *107* (9), 097204.
63. Tapp, K., *Differential Geometry of Curves and Surfaces*. 1 ed.; Springer Cham: 2016; p VIII, 366.
64. Zhang, Y.; Liu, J.; Dong, Y.; Wu, S.; Zhang, J.; Wang, J.; Lu, J.; Rückriegel, A.; Wang, H.; Duine, R.; Yu, H.; Luo, Z.; Shen, K.; Zhang, J., Strain-Driven Dzyaloshinskii-Moriya Interaction for Room-Temperature Magnetic Skyrmions. *Physical Review Letters* **2021,** *127* (11), 117204.
65. Davydenko, A. V.; Kozlov, A. G.; Chernousov, N. N.; Turpak, A. A.; Ermakov, K. S.; Stebliy, M. E.; Letushev, M. E.; Sadovnikov, A. V.; Golikov, A. P.; Ognev, A. V.; Shiota, Y.; Ono, T.; Samardak, A. S., Giant Asymmetry of Domain Walls Propagation in Pd/Co/Pd(111) Epitaxial Structures with Different Interface Roughness. *ACS Applied Electronic Materials* **2024,** *6* (2), 1094-1103.
66. Samardak, A. S.; Davydenko, A. V.; Kolesnikov, A. G.; Samardak, A. Y.; Kozlov, A. G.; Pal, B.; Ognev, A. V.; Sadovnikov, A. V.; Nikitov, S. A.; Gerasimenko, A. V.; Cha, I. H.; Kim, Y. J.; Kim, G. W.; Tretiakov, O. A.; Kim, Y. K., Enhancement of perpendicular magnetic anisotropy and Dzyaloshinskii–Moriya interaction in thin ferromagnetic films by atomic-scale modulation of interfaces. *NPG Asia Materials* **2020,** *12* (1), 51.
67. Kozlov, A. G.; Davydenko, A. V.; Afremov, L. L.; Iliushin, I. G.; Kharitonov, V. N.; Mushtuk, P. S.; Tarasov, E. V.; Turpak, A. A.; Shishelov, A. F.; Chernousov, N. N.; Letushev, M. E.; Sadovnikov, A. V.; Khutieva, A. B.; Ognev, A. V.; Samardak, A. S., Effects of Interfacial Nanoengineering through an Artificial Oxidation of Epitaxial Pd/Co Ultrathin Films on Perpendicular Magnetic Anisotropy and the Dzyaloshinskii–Moriya Interaction. *ACS Applied Electronic Materials* **2024,** *6* (7), 4928-4938.

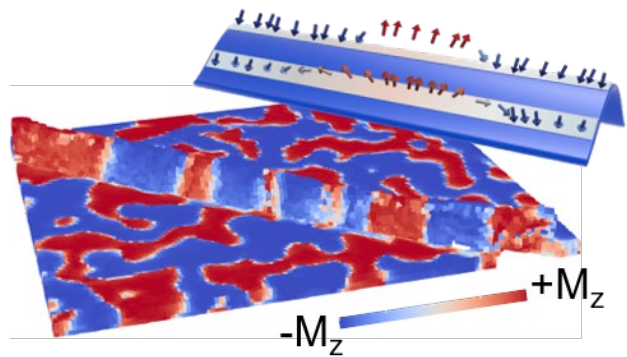

**TOC figure:** X-ray nanotomography of a curved magnetic thin film with perpendicular anisotropy reveals curvature-induced modifications in the magnetic configuration. The curvature drives the magnetization toward the local surface normal and significantly alters the domain alignment. Furthermore, curvature-induced Dzyaloshinskii–Moriya interaction (DMI) enhances the stability of Néel-type domain walls.